\documentclass[aps,pre,reprint,floatfix,superscriptaddress,showpacs]{revtex4-1}
\usepackage{graphicx}
\usepackage[alsoload=synchem]{siunitx}
\DeclareSIUnit{\cal}{cal}
\DeclareSIUnit{\Newton}{N}
\usepackage{amsmath,amssymb}
\RequirePackage{lineno} 
\usepackage[nodayofweek]{datetime}
\newdateformat{myDate}{\THEDAY\ \monthname[\THEMONTH] \THEYEAR}
\usepackage{hyperref}
\hypersetup{colorlinks, citecolor={blue},linkcolor={red}, urlcolor={blue},
pdftitle={duplex rupture},
pdfauthor={Majid Mosayebi}, pdfdisplaydoctitle}

\begin{document}

\title{Force-induced rupture of a DNA duplex}

\author{Majid Mosayebi}
\email{majid.mosayebi@chem.ox.ac.uk}
\affiliation{Physical and Theoretical Chemistry Laboratory, Department of Chemistry, University of Oxford, South Parks Road, Oxford, OX1 3QZ, United Kingdom}
\author{Ard A. Louis}
\affiliation{Rudolf Peierls Centre for Theoretical Physics, 1 Keble Road, Oxford OX1 3NP, United Kingdom}
\author{Jonathan P.~K.~Doye}
\affiliation{Physical and Theoretical Chemistry Laboratory, Department of Chemistry, University of Oxford, South Parks Road, Oxford, OX1 3QZ, United Kingdom}
\author{Thomas E. Ouldridge}
\email{t.ouldridge@imperial.ac.uk}
\affiliation{Rudolf Peierls Centre for Theoretical Physics, 1 Keble Road, Oxford OX1 3NP, United Kingdom}
\affiliation{Department of Mathematics, Imperial College, 180 Queen's Gate, London SW7 2AZ, United Kingdom}


\begin{abstract}
The rupture of double-stranded DNA under stress is a key process in biophysics and nanotechnology. In this article we consider the shear-induced rupture of short DNA duplexes, a system that has been given new importance by recently designed force sensors and nanotechnological devices. We argue that rupture must be understood as an activated process, where the duplex state is metastable and the strands will separate in a finite time that depends on the duplex length and the force applied. Thus, the critical shearing force required to rupture a duplex within a given experiment depends strongly on the time scale of observation. We use simple models of DNA to demonstrate that this approach naturally captures the experimentally observed dependence of the critical force on duplex length for a given observation time. In particular, the critical force is zero for the shortest duplexes, before rising sharply and then plateauing in the long length limit. The prevailing approach, based on identifying when the presence of each additional base pair within the duplex is thermodynamically unfavorable rather than allowing for metastability, does not predict a time-scale-dependent critical force and does not naturally incorporate a critical force of zero for the shortest duplexes. Additionally, motivated by a recently proposed force sensor, we investigate application of stress to a duplex in a mixed mode that interpolates between shearing and unzipping. As with pure shearing, the critical force depends on the time scale of observation; at a fixed time scale and duplex length, the critical force exhibits a sigmoidal dependence on the fraction of the duplex that is subject to shearing.\end{abstract}

\maketitle

\section{Introduction}
Understanding the properties of stress-induced unwinding of a DNA duplex is crucial in analyzing many biophysical systems. Within the cell, enzymes exert large forces on DNA;  for example in replication or transcription \cite{helicase_nar}. In the growing field of DNA nanotechnology, duplexes within self-assembled structures and devices can be deliberately subjected to substantial stresses \cite{Dietz2009,Liedl2010,Han2011,tom_walker,Loh2014} to enhance the design possibilities. Recently, force-induced rupture of DNA has even been pioneered as a mechanism for measuring forces \cite{Ho_DNAsensor, Wang_TGT,Blakely_force_sensor,Zhang_force_sensor}; for example, Wang and Ha used duplex rupture to estimate the force that a mechano-sensitive receptor in a cell experiences from a surface \cite{Wang_TGT}. The importance of the response of DNA to stress and recent progress in single-molecule force-spectroscopy techniques \cite{Lee_science, Bustamante_review} have led to many detailed investigations of stress-induced duplex disruption \cite{Pope2000, Strunz_forcespec, Sattin2004, vanMameren2009, Huguet2010, Salerno2012,  Tempestini2013, Schumakovitch_T_rupture, Grange2001, Neuert2007, Hatch_pulling, Danilowicz2003, Danilowicz_unzipping}. Simultaneously, theoretical models have been developed to predict and explain the experimental results \cite{Strunz_forcespec, Pope2000, Neuert2007, Danilowicz2003, Huguet2010, Hatch_pulling, Marenduzzo_unzipping, deGennes_ladder_model, Chakrabarti_shearing, Lubensky_unzipping, Cocco_unzipping, Cocco2004, Singh_lattice_rupture, Kumar_priodic_force, Kumar_PhysRep, Prakash2011, Mishra2011}.

\begin{figure}
  \centering
  \includegraphics[width=0.6\linewidth]{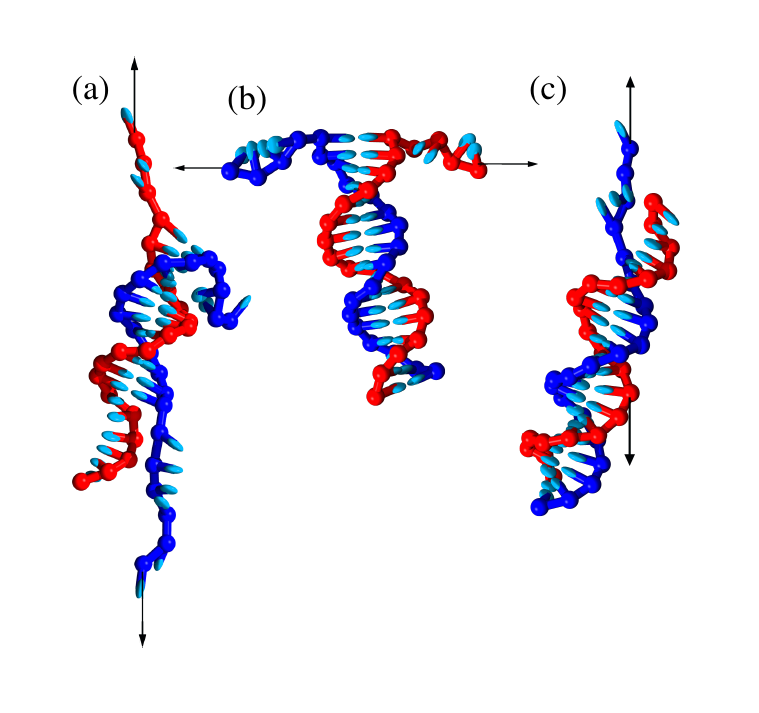}
 \caption{ Applying stress to short DNA molecules. Depending on where the two anti-parallel forces are applied, DNA can be subject to stress in different modes: (a) shearing mode, (b) unzipping mode and (c) mixed mode. In the shearing mode, the maximum extension per base pair gained from disrupting the duplex prior to complete strand separation ($\sim 0.3$\,nm) is much less than in the unzipping mode  ($\sim 1.3$\,nm), meaning that duplexes tend to unzip at lower forces than they shear \cite{Hatch_pulling,Huguet2010}.
 }
  \label{fig:modes}
\end{figure}

In this article we study the physics underlying shearing of short DNA duplexes, illustrated in Fig.\,\ref{fig:modes}\,(a). We argue that the currently prevailing theory \cite{Hatch_pulling, deGennes_ladder_model, Chakrabarti_shearing,Prakash2011,Mishra2011} fundamentally misinterprets the behavior of the critical rupture force as a function of duplex length.  We first frame our discussion with a toy model to highlight the basic physics, and review previous approaches to understanding the problem. We then explore the system in more depth with a detailed coarse-grained DNA model (oxDNA). Our interpretation emphasizes the finite time scale of experimental observation, and predicts a critical force that depends on this time scale as well as duplex length. OxDNA also allows us to explore the mechanosensor of Wang and Ha \cite{Wang_TGT}, which involves a generalization of shearing.

\subsection{Background theory: Basic models of shearing}
\begin{figure*}
  \centering
  \includegraphics[width=\linewidth]{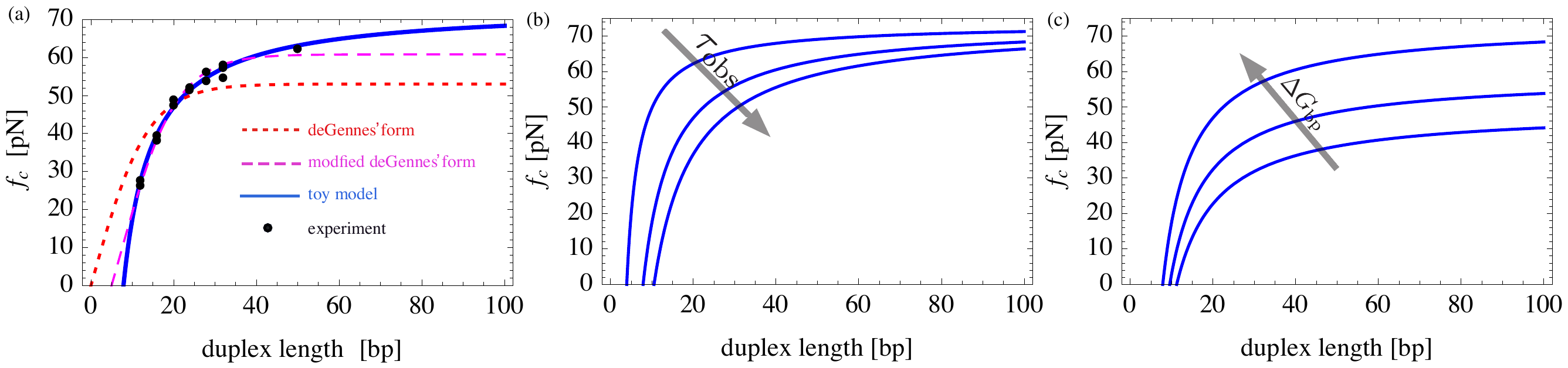}
 \caption{Behavior of the critical shearing force. (a) Critical force $f_c$ as a function of duplex length as observed in experiment (circles \cite{Hatch_pulling}) and predicted by basic models of shearing. The toy model is shown as a solid line; a two parameter fit to the experimental data was performed, yielding $\Delta G^\ddagger_{\tau_{\rm obs}}/\delta=74.3$\,\SI{}{\kilo\cal\per\mole\per\nm} and $\Delta G_{\rm bp}/\delta = 10.9$\,\SI{}{\kilo\cal\per\mole\per\nm}. The results are compared to those of deGennes' original form \cite{deGennes_ladder_model} (dotted line) and the modified deGennes' form \cite{Hatch_pulling} (dashed line), in both cases using the parameters obtained by Hatch  {\it et al.}\:for fitting the modified form to the experimental data \cite{Hatch_pulling}. (b) Illustration of the dependence of the critical rupture force $f_c$ on $\Delta G^\ddagger_{\tau_{\rm obs}}$ and hence the observation time as predicted by the toy model. $\Delta G^\ddagger_{\tau_{\rm obs}}$, and hence $\tau_{\rm obs}$, increases from top to bottom ($\Delta G^\ddagger_{\tau_{\rm obs}} \in \{4.4, 10.4, 14.4\}\,$\SI{}{\kilo\cal\per\mole}), and $\Delta G_{\rm bp}=1.52$\,\SI{}{\kilo\cal\per\mole}. (c) Dependence of the rupture force on base pair stability $\Delta G_{\rm bp}$, as predicted by the toy model. $\Delta G_{\rm bp}$ decreases from top to bottom ($\Delta G_{\rm bp} \in \{1.52, 1.22, 1.02\}\,$\SI{}{\kilo\cal\per\mole}), and $\Delta G^\ddagger_{\tau_{\rm obs}}=10.4\,$\SI{}{\kilo\cal\per\mole}. In (b) and (c), $\delta = 0.14$\,\SI{}{\nm}.
 }
  \label{fig:toymodel}
\end{figure*}

Shearing of a duplex involves applying antiparallel forces to either the opposite $3^\prime$-$3^\prime$ or $5^\prime$-$5^\prime$ ends of the bound duplex, as shown in Fig.\,\ref{fig:modes}(a). Such a setup has been explored experimentally \cite{Hatch_pulling, Wang_TGT, Pope2000, Strunz_forcespec, Sattin2004, Schumakovitch_T_rupture, Grange2001,Neuert2007} and theoretically \cite{Hatch_pulling, deGennes_ladder_model, Chakrabarti_shearing, Singh_lattice_rupture, Prakash2011, Mishra2011, Strunz_forcespec, Neuert2007, Pope2000}. In early experiments \cite{Pope2000, Strunz_forcespec, Sattin2004, Schumakovitch_T_rupture, Grange2001,Neuert2007}, loads were increased dynamically; more recently, Hatch {\it et al.}\:were able to use a constant force \cite{Hatch_pulling}. The critical force $f_c$ above which the duplex ruptures is a key observable; Hatch {\it et al.}\:measured the shearing force for duplexes from length 12 to 50 bp (base pairs), and found that it increased with duplex length at short lengths before plateauing in the long length limit \cite{Hatch_pulling} (Fig.\,\ref{fig:toymodel}\,(a)). Clearly, for any non-zero force, the stable state involves widely separated strands. At finite temperature, therefore, any duplex will rupture given sufficient time. Thus, the critical force should be understood as the force required to rupture a duplex within a given experimental time scale, and it is fundamentally important to view rupture under shearing in the context of time scales of experimental observation. For systems in which load is dynamically applied, the time scale over which the force changes is an additional complicating factor. We will restrict our discussion to the constant force case.

At room temperature, opening of a duplex is widely believed to be a thermally activated process \cite{Saenger1984,Morrison1993,Cocco_unzipping,eyal_hairpin,tom_hybridization}. The system must pass through an unfavorable transition state in which the strands are attached by a small number of base pairs; one would expect the rate to be exponentially suppressed by the free energy cost $\Delta G^\ddagger$ of reaching this state from the fully formed duplex. Assuming a typical free-energy cost of base-pair disruption of $\Delta G_{\rm bp}$ \cite{SantaLucia2004},  $\Delta G^\ddagger = N \Delta G_{\rm bp} - \Delta G_0$ where $N$ is the duplex length and $\Delta G_0$ is an $N$-independent constant that incorporates, for example, the typical number of base pairs in a transition state. To a first approximation, the effect of a constant shearing force $f$ is to modulate $\Delta G^\ddagger$ as the transition state and duplex state have different extensibilities.
\begin{equation}
\Delta G^\ddagger(f) = N \Delta G_{\rm bp} - \Delta G_0 + \int_0^f  \big( x_{\rm d} (f^\prime) - x_{\rm t} (f^\prime) \big)  {\rm d} f^\prime,
\label{eq:dGdagger}
\end{equation}
in which $ x_{\rm d} (f)$ is the average extension of the fully-formed duplex under force $f$, and  $x_{\rm t} (f)$ is the extension of the transition state. Evaluating the integral in Eq.\,\ref{eq:dGdagger} necessitates a detailed model of DNA, but physical insight can be obtained through the crude approximation of $ \int_0^f  (x_{\rm d} (f^\prime) - x_{\rm t} (f^\prime))  {\rm d} f^\prime = - (N \delta - \delta_0)f$  , in which $\delta$ is a positive constant representing the degree per base pair to which the disrupted duplex will tend to have a larger extension due to its single-stranded state. $\delta_0$, like $\Delta G_0$, accounts for the number of base pairs in the transition state.

Given the above description, a simple model for the rate $1/\tau(f)$ of duplex rupture as a function of duplex length and force would be 
\begin{equation}
\begin{array}{c}
\dfrac{1}{ \tau(f)} =k_0 \exp(-(\Delta G^\ddagger(f))/kT) \vspace{2mm}\\
= k_0 \exp(-(N \Delta G_{\rm bp} - \Delta G_0 - (N \delta - \delta_0 )f )/kT),
\end{array}
\label{eq:Arrhenius}
\end{equation}
in which $k_0$ is an  unknown rate constant. Thus for rupture to occur on a time scale of experimental observation $\tau_{\rm obs}$, we require $\Delta G^\ddagger(f)) \lesssim \Delta G_{\tau_{\rm obs}}^\ddagger = kT \ln (\tau_{\rm obs} k_0)  $. This expression defines a critical force below which duplexes will not tend to rupture during the experiment, and above which rupture is typical:
\begin{equation}
f_c(N) = \frac{N \Delta G_{\rm bp}}{N \delta - \delta_0 } - \frac{\Delta G_{\tau_{\rm obs}}^\ddagger+\Delta G_0}{ N \delta - \delta_0 }.
\label{eq:toy_prediction}
\end{equation}
At large $N$, $f_c(N) \approx \Delta G_{\rm bp} /\delta$ and is $N$-independent. As $N$ drops, the importance of the second term on the right-hand side grows and $f_c(N)$ therefore decreases, reaching zero when $N\Delta G_{\rm bp} - \Delta G_0 = \Delta G_{\tau_{\rm obs}}^\ddagger$. Thus our simple model predicts that duplexes below a certain length will dissociate spontaneously on an experimental time scale; above this length the critical force required for rupture will rise from 0 and eventually plateau in the long length limit. Although the toy model is crude, the underlying physics is reasonable: the shearing stress acts to reduce the stability per base pair of the duplex state, allowing duplexes to rupture within a smaller observation time $\tau_{\rm obs}$ than at zero force. Long duplexes require a higher force than short ones, since if more base pairs are present then each individual base pair must be weakened further to result in the same barrier height $\Delta G^\ddagger_{\tau_{\rm obs}}$.  At high enough forces $f \delta > \Delta G_{\rm bp}$, breaking a base pair is thermodynamically favorable and so even long duplexes will rupture rapidly, explaining the plateau in $f_c$ at large $N$.

The behavior of Eq.\,\ref{eq:toy_prediction} is illustrated in Fig.\,\ref{fig:toymodel}\,(a), in which the toy model is fitted to the experimental data of Hatch {\it et al.}\:\cite{Hatch_pulling}. For simplicity, we assume that the transition state is a single base-pair state; we thus set $\delta_0 = \delta$ and $\Delta G_0 = \Delta G_{\rm bp}$ and fit $\Delta G_{\rm bp}/\delta$ and $\Delta G_{\tau_{\rm obs}}^\ddagger/\delta$ to the data. The observed behavior is clearly consistent with experiment. Further, the fitting parameters are reasonable given the crudeness of the model. If $\delta \approx 0.15$\,nm, $\Delta G_{\rm bp} \approx 1.5$ \SI{}{\kilo\cal\per\mole}, similar to values reported by Ref. \cite{SantaLucia2004}, and the barrier height $\Delta G^\ddagger_{\tau_{\rm obs}}/\Delta G_{\rm bp} \approx 6{\rm{-}}7$, implying that duplexes of $7$-$8$\,bp can dissociate within the observation time (1\,s),  which is not unreasonable \cite{Morrison1993,Zhang_disp_2009, eyal_hairpin}. Importantly, as shown in Fig.\,\ref{fig:toymodel}\,(b), the shape of $f_c(N)$ depends on $\Delta G_{\tau_{\rm obs}}^\ddagger$. From Eq.\,\ref{eq:toy_prediction}, $f_c$ depends linearly on $\Delta G_{\tau_{\rm obs}}^\ddagger$ and hence (through Eq.\,\ref{eq:Arrhenius}) it  decreases logarithmically with increasing $\tau_{\rm obs}$. This dependence is physically reasonable; for example, the maximal length of duplex that will spontaneously dissociate (rupture at $f=0$) in a time scale of milliseconds is clearly much less than the maximal length that will spontaneously dissociate in a time scale of years. It is also consistent with the fact that when stress is increased over time, rather than held constant, the duplex rupture force increases with the rate at which the force is applied \cite{Pope2000, Strunz_forcespec, Sattin2004, Schumakovitch_T_rupture, Grange2001, Neuert2007}. Nevertheless, regardless of how long the observation time, for infinite length chains the critical force will still plateau at $f_c(\infty)=\Delta G_{\rm bp}/\delta$; this limiting behavior is simply reached more slowly with $N$ for larger $\tau_{\rm obs}$.  

In Fig.\,\ref{fig:toymodel}\,(c) we show the effect of varying $\Delta G_{\rm bp}$  on $f_c$.  Reducing this base-pair stability leads to a lower plateau height $f_c(\infty)$ and  sets the lower cutoff length below which $f_c = 0$ to larger $N$.

Broadly similar reasoning to that underlying the toy model has been applied to the unzipping of DNA (in which two strands are pulled apart from the same end of the duplex -- see Fig.\,\ref{fig:modes}\,(b)) \cite{Cocco_unzipping}. Experiments in which force is applied dynamically have also been analyzed in terms of models of activated processes \cite{Strunz_forcespec, Pope2000, Neuert2007}. Despite this history, the physical principles highlighted by the toy model have not been widely applied to understand the dependence of the critical shearing force on duplex length \cite{Hatch_pulling, deGennes_ladder_model, Chakrabarti_shearing,Prakash2011,Mishra2011}. The alternative reasoning originated from deGennes, who modelled DNA as a ladder with springs connecting both neighbors within a strand and bases paired by interstrand hydrogen bonding \cite{deGennes_ladder_model}. He calculated the mechanical equilibrium of this system under applied shearing stress, and posited that an individual base-pair spring with extension above a certain critical value would rupture. DeGennes showed that, for longer strands, the shearing stress can be spread out over several base pairs in mechanical equilibrium, reducing the strain at the duplex ends and allowing the duplex to withstand higher stresses than a single isolated base pair. His result for the critical force is \cite{deGennes_ladder_model}
\begin{equation}
f_c = 2 f_1 \chi^{-1} \tanh \left( \chi N/2 \right),
\label{eq:dG_prediction}
\end{equation}
in which $f_1$ is the force required to rupture a single base pair and $\chi = \sqrt{2R/Q}$ is a function of the spring constant between neighbors in a strand, $Q$, and the spring constant between base pairs $R$. Similarly to Eq.\,\ref{eq:toy_prediction}, Eq.\,\ref{eq:dG_prediction} also has an $f_c$ that increases with $N$ and plateaus in the long length limit. The underlying physics is, however, entirely different. The deGennes model assumes that rupture can only occur when each base pair is individually unstable; it does not allow for metastability and a finite time scale for rupture during which the system can break many base pairs to climb over a free-energy barrier. The shape of the curve is governed by $\chi$, a property of the duplex state, rather than the relative properties of single-stranded and duplex DNA. A number of groups have explored extensions and improvements to the deGennes' model \cite{Hatch_pulling, deGennes_ladder_model, Chakrabarti_shearing,Prakash2011}, but their approaches remain fundamentally based on the idea that the dependence of $f_c$ on $N$  can be explained by the degree to which longer duplexes can spread mechanical stress over multiple base pairs, rather than because longer duplexes need a larger force to reduce the free-energy barrier opposing dissociation to a low enough value. 

As well as being based on fundamentally different physics, the two approaches also make qualitatively distinct predictions. Most obviously, Eq.\,\ref{eq:dG_prediction} predicts a finite rupture force for $N=1$, whereas Eq.\,\ref{eq:toy_prediction} predicts $f_c > 0$ only for $N> N_0 = (\Delta G_{\tau_{\rm obs}}^\ddagger + \Delta G_0)/\Delta G_{\rm bp}$. Further, Eq.\,\ref{eq:toy_prediction}  predicts that $f_c(N)$ depends logarithmically on the observation time $\tau_{\rm obs}$ through its linear dependence on $\Delta G_{\tau_{\rm obs}}^\ddagger $; by contrast, the deGennes theory does not include $\tau_{\rm obs}$. 

The experimental data from \cite{Hatch_pulling} in Fig.\,\ref{fig:toymodel}\,(a) are suggestive of $f_c > 0$ only above a certain value of $N= N_0 \sim 5-8$\,bp, as in fact was recognized by the authors. Nonetheless, they attempted to fit the deGennes model to their data \cite{Hatch_pulling}, modifying Eq.\,\ref{eq:dG_prediction} to
\begin{equation}
f_c = 2 f_1  \chi^{-1} \tanh \left( \chi (N-N_{\rm open})/2 \right) +2 f_1
\label{eq:hatch_prediction}
\end{equation}
which includes a new parameter $N_{\rm open}$ and an additional term  $2 f_1$.
The first parameter was set to $N_{\rm open} = 7$, in order to generate a finite $N_0$.  They argued for a reduced effective duplex length $N-7$\,bp on the grounds that the end base pairs are always broken at room temperature. However, such an approach is inconsistent with the thermodynamics of DNA as currently understood \cite{SantaLucia2004}; for example, hairpins with stems of three or four base pairs can be stable relative to the unfolded state at room temperature \cite{SantaLucia2004,Gao2006}. Additionally, this approach does not include a dependence on $\tau_{\rm obs}$ as we argued above; the curve is identical for experiments that take milliseconds and experiments that take years. Finally, the value of $\chi$ obtained from their fit, 0.147, is unphysically low. $\chi = 0.147$ would imply $Q/R=92.5$, {\it i.e.} that stretching hydrogen bonds between base pairs is almost two orders of magnitude easier than extending the distance between stacked bases in a duplex. Such an unreasonable value, necessary to force the curve to reach large $N$ values before plateauing, provides further evidence that the deGennes theory does not explain the shape of $f_c(N)$. We note in passing that the reason for the term $2 f_1=7.8$\,\SI{}{\pico \Newton} term in Eq.\,\ref{eq:hatch_prediction}  as compared to Eq.\,\ref{eq:dG_prediction} is unclear; this term shifts the curve slightly, but does not strongly influence the above discussion.


Although the toy model is useful in providing understanding, it is extremely crude and neglects a number of potentially important effects. Most significantly, taking $ \int_0^f  (x_{\rm d} (f^\prime) - x_{\rm t} (f^\prime) ) {\rm d} f^\prime = - (N \delta - \delta_0)f$ in Eq.\,\ref{eq:dGdagger} is a very strong approximation. In this work we therefore use oxDNA \cite{tom_model_jcp, tom_thesis,petr_seq_dep}, a nucleotide level coarse-grained model of DNA that captures both the elastic behavior of double and single strands, as well as the basic physics of bonding.  It should therefore be able to describe in detail the effects of the shearing forces on a duplex, and capture metastable states,  which we will argue are critical in order to obtain an $f_c(N)$ that is consistent with experiment. We note that oxDNA has the advantages of representing DNA structure and mechanics more accurately than simpler statistical models such as those of Refs. \cite{deGennes_ladder_model} and  \cite{Chakrabarti_shearing}. Further, oxDNA is not derived with either the assumptions of the toy model or of the deGennes' model and its derivatives in mind. Rather, we use it to explore whether the findings of the toy model are robust to a more accurate model of DNA thermodynamics and mechanics, and whether effects analogous to those identified by deGennes are likely to play a significant role.

\section{Methods}
In oxDNA, each nucleotide is treated as a rigid body \cite{tom_model_jcp, tom_thesis,petr_seq_dep}. 
OxDNA nucleotides interact through potentials designed to mimic hydrogen-bonding, stacking, chain connectivity and excluded volume interactions; these interactions combine to allow the formation of right-handed double helices between complementary strands at low temperatures. OxDNA incorporates physically reasonable representations of the thermodynamics, mechanics and structure of single-stranded and duplex DNA \cite{tom_model_jcp, tom_thesis} --- the key ingredients whose interplay is central to this system. 
The model has been shown to reproduce important aspects of basic processes such as hybridization \cite{tom_hybridization}, toehold mediated strand displacement \cite{tom_toehold} and hairpin formation \cite{hairpin_stacking}. It has also been successfully applied to explore stress-induced transitions \cite{matek2012, flavio_overstretching,plectoneme}. In this paper we use a sequence-averaged parameterization of oxDNA \cite{tom_model_jcp,tom_thesis}, which is ideal for identifying generic trends.

Since direct measurements of the rupture kinetics are very demanding, we estimate rates indirectly from the free-energy profiles of stressed duplexes.
We simulate the model using the virtual move Monte Carlo (VMMC) algorithm of Whitelam and Geissler (the variant in the appendix of Ref. \cite{Whitelam2009}). Shear stress is generated  by applying two anti-parallel forces of magnitude $f$ to the center of mass of the nucleotides at the $3^\prime$ end of each strand (which are situated at opposite ends of the duplex). We measure the free-energy profile as a function of the number of base pairs between the two strands (a base pair being defined by a hydrogen-bonding energy of less than $−0.596$\,\SI{}{\kilo\cal\per\mole}. A cut-off which is around $15\%$ of the typical hydrogen-bonding energy in the model), using umbrella sampling \cite{us} to facilitate the measurement of less favorable states. We also use umbrella sampling to prohibit the strands from separating (once separated, the strands would never rebind), and for simplicity we only include hydrogen-bonding interactions between native base pairs (those present in the fully-formed duplex). It was shown \cite{tom_hybridization} that in oxDNA non-native base pairs have only a minor effect for non-repetitive sequences on the hybridization transition in the absence of force; the consequences of non-native base pairs for shearing induced rupture will be studied in detail elsewhere \cite{rupture_jcp}.

We obtain the free-energy profiles at different force values, separated by 2 to 5 \SI{}{\pico \Newton} and spanning our range of interest, from umbrella sampling simulations at $T_0=23$\SI{}{\degreeCelsius}. The umbrella weights were adjusted iteratively to have a uniform sampling as a function of the number base pairs in the duplex. The starting configuration at each stage was chosen to be the final configuration from the previous iteration. At each force value, we ran 3 to 5 independent simulations, each with approximately $10^9$ VMMC steps per particle. The resulting free energies from independent simulations agreed, confirming that our free-energy profiles are indeed converged. The value of the rupture force $f_c$ for a given $\Delta G^\ddagger_{\tau_{\rm obs}}$ is then obtained via fitting the rupture force versus barrier height data to an interpolating piecewise cubic Hermite polynomial $\mathcal{P}$; $f_c=\mathcal{P}(\Delta G^\ddagger_{\tau_{\rm obs}})$. We determine the rupture force at other temperatures $(T \neq T_0)$, from extrapolated free-energy profiles obtained using single histogram reweighting, based on the method of Ferrenberg and Swensden \cite{Ferrenberg_extrap}. Note that the accuracy of the extrapolated free energies decreases when $T$ is far from $T_0$. Therefore, we restricted our considerations to temperatures where $|T-T_0| \leq 12$  \SI{}{\degreeCelsius}.

\begin{figure*}
  \centering
  \includegraphics[width=\linewidth]{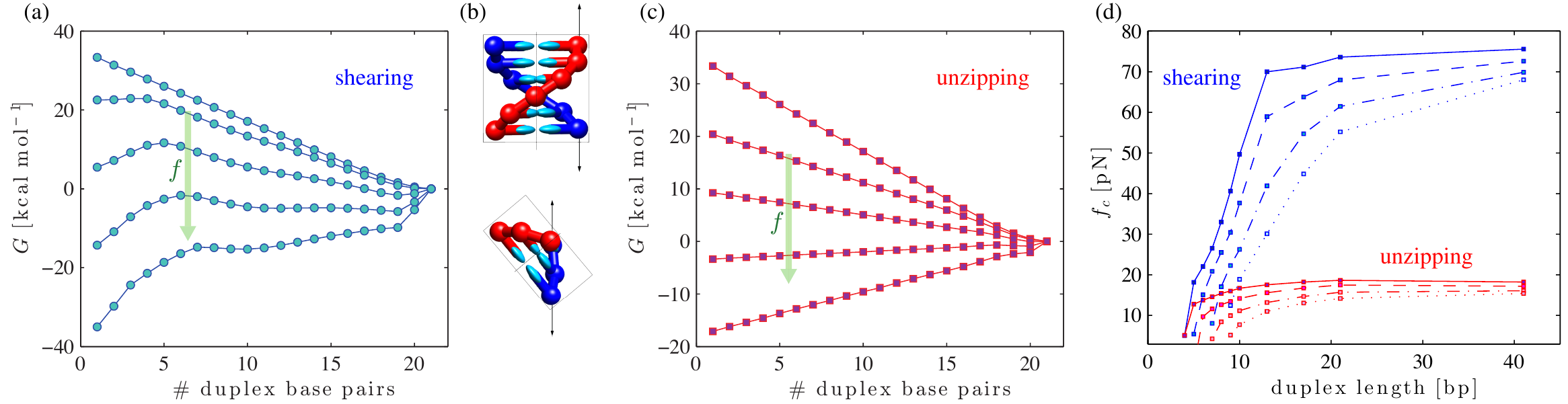}
 \caption{
 Shearing and unzipping as represented by oxDNA. (a)
 Free-energy profiles of the bound duplex of length 21 base pairs at $T=23$\SI{}{\degreeCelsius} in the shearing mode. The pulling force increases from top to bottom: $f^{\rm shear} \in \{0, 30, 50, 70, 90 \}$\SI{}{\pico \Newton}. (b) Rotation of the helix to maximize extension without disrupting structure. A 6-base-pair helix has its maximum extension when its helical axis is aligned with the stress; a 3-base-pair helix can rotate to increase its extension along the force. (c)  Free-energy profiles of the bound duplex of length 21 base pairs at $T=23$\SI{}{\degreeCelsius} in the unzipping mode. The pulling force increases from top to bottom:  $f^{\rm unzip} \in \{0, 10, 15, 20, 25\}$\SI{}{\pico \Newton}. (d) Rupture force as a function of duplex length in shearing and unzipping modes.  The critical free-energy barrier height increases from top to bottom ($\Delta G^\ddagger_{\tau_{\rm obs}} \in \{3, 5, 8, 11\}$ \SI{}{\kilo\cal\per\mole}). Larger $\Delta G^\ddagger_{\tau_{\rm obs}}$ corresponds to longer experimental measuring times.
 }
 \label{fig:free-energy}
\end{figure*}

\section{Results and Discussion}
\subsection{Pure shearing and unzipping}
We measure free-energy profiles of shearing for a range of forces and duplex lengths; typical profiles are shown in Fig.\,\ref{fig:free-energy}\,(a). At low forces, formation of each additional base pair is thermodynamically favorable but as the shear force is increased the slope of the profile becomes shallower and the duplex becomes less stable. On top of these overall trends, we see features that reflect the helical geometry of the stressed duplex. A duplex that has intact base pairs equal to a half turn ({\textit i.e.} approximately 6 to 7 base pairs) has its maximum extension when the duplex axis is aligned with the force. When such a duplex loses or gains base pairs it can rotate its helical axis away from the direction of the force, allowing a greater extension per base pair (see Fig.\,\ref{fig:free-energy}\,(b)). The result is a non-monotonic dependence of the free energy on the number of duplex base pairs; the effect is strongest for short duplex sections, and explains the peak in free energy that appears at approximately 6 to 7 base pairs as the force increases. Such geometrical effects are absent in deGennes' ladder model \cite{deGennes_ladder_model} and in most subsequent developments of that approach, although a similar effect in the context of a 2D ladder model was considered  by  Chakrabarti and Nelson \cite{Chakrabarti_shearing}.

For comparison, we also measure the profile for unzipping (when the two strands are pulled apart from the same end of the duplex, see Fig.\,\ref{fig:modes}\,(b)). Typical behavior as a function of $f$ is shown in Fig.\,\ref{fig:free-energy}\,(c). As with shearing, base pairs become less stable as the force is increased. Two differences with shearing are clear. Firstly, the forces required to reduce stability are much lower; this is because a single unzipped base pair increases the extension of the DNA along the applied force much more than a single sheared base pair (this is evident in Fig.\,\ref{fig:modes}). Secondly, the non-monotonicity observed for shearing is absent for this geometry; regardless of the number of base pairs present it is always favorable to keep the helix axis perpendicular to the unzipping force. 

OxDNA clearly shows a more complex dependence of free energies on $f$ than incorporated into the simple model. Nonetheless, we can identify a barrier height $\Delta G^\ddagger_{\rm sim}$ in each profile as the free-energy difference between the (local) maximum in the profile with the smallest number of base pairs, and the lowest (local) minimum in the profile that has more base pairs than the maximum. We take these configurations to be proxies for the transition state and the duplex state, respectively, and assume that the dissociation rate is given by  $1/\tau(f) \propto \exp(-(\Delta G^\ddagger_{\rm sim})/kT)$. Previous studies with oxDNA in the absence of stress have shown that attachment rates of duplexes are only weakly dependent on overall duplex stability \cite{tom_toehold, tom_hybridization}; this is consistent with the assumption that the detachment rate is largely determined by $\exp(-(\Delta G^\ddagger_{\rm sim})/kT)$, as this barrier is the dominant factor in duplex stability. 

\begin{figure}
  \centering
  \includegraphics[width=0.74\linewidth]{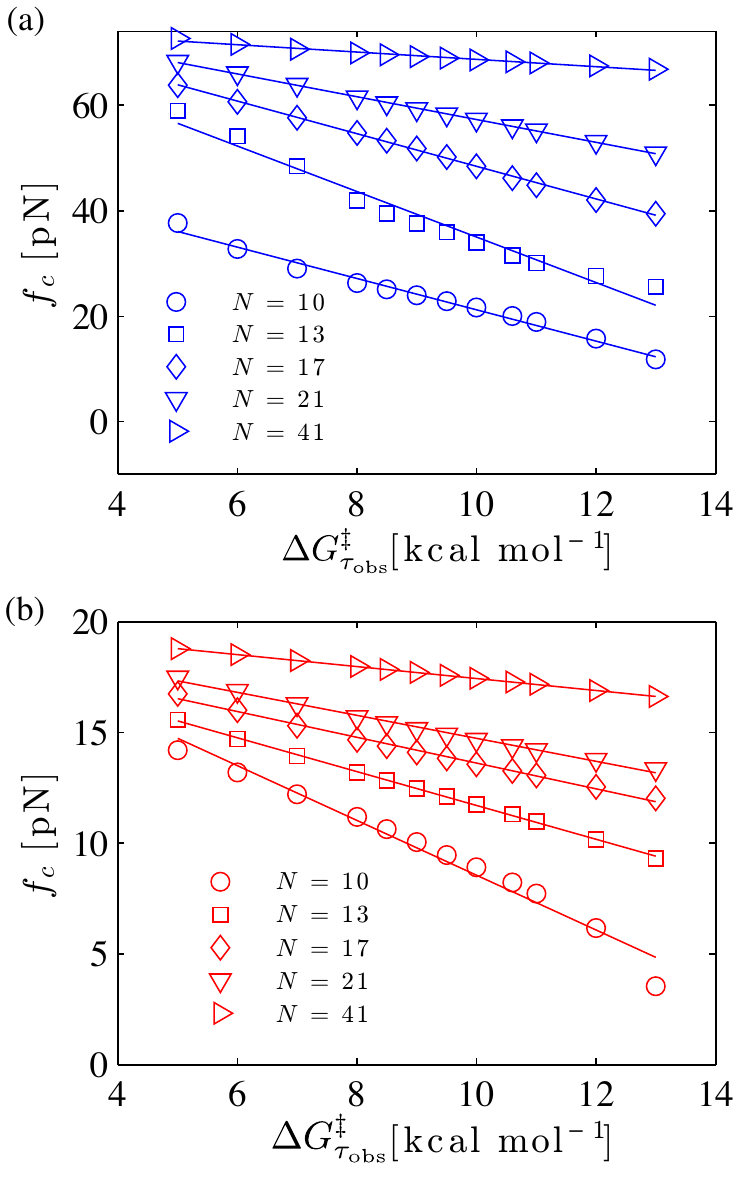}
 \caption{The dependence of $f_c$ on $\Delta G^\ddagger_{\tau_{\rm obs}}$ at fixed $N$ as estimated by oxDNA.
(a) Shearing mode. (b) Unzipping mode. The apparent non-monotonicity in the fitted gradient that is present for shearing but absent for unzipping is due to the number of base pairs in the transition state changing with applied force.  
  }
  \label{fig:fc_as_fn_G}
\end{figure}

As with the toy model, we can define a rupture force $f_c$ that is the force required to reduce $\Delta G^\ddagger_{\rm sim}$ to a certain value $\Delta G^\ddagger_{\tau_{\rm obs}}$,  set by the time scale of experimental observation. We plot critical rupture forces for unzipping and shearing as a function of duplex length $N$ in Fig.\,\ref{fig:free-energy}\,(d) for a range of $\Delta G^\ddagger_{\tau_{\rm obs}}$ values.  Despite the additional complexity of oxDNA, the results are broadly consistent with the toy model; in particular, the dependence of $f_c(N)$ on the observation time is similar. In Eq.\,\ref{eq:toy_prediction} the toy model predicts a linear dependence of $f_c(N)$ on $\Delta G^\ddagger_{\tau_{\rm obs}}$ (and hence a logarithmic dependence on $\tau_{\rm obs}$) at fixed $N$; Fig.\ref{fig:fc_as_fn_G}\,(a) shows that oxDNA is in broad agreement. The agreement is not perfect, however, and in particular Eq.\,\ref{eq:toy_prediction} predicts that the magnitude of the gradient of $f_c$ with $\Delta G^\ddagger_{\tau_{\rm obs}}$ should decrease monotonically as $N$ increases; this is not apparent in our fits as the gradient for $N=10$ is shallower than for $N=13$. In fact, this is due to non-linearities in the dependence of $f_c$ on $\Delta G^\ddagger_{\tau_{\rm obs}}$ (when viewed at constant $f_c$, the $N=10$ curve is actually steeper). This non-linearity is inconsistent with the toy model, which assumes that the properties of the transition state do not change with force. It is clear from Fig.\,\ref{fig:free-energy}\,(a), however, the transition state moves to higher numbers of base pairs as the force increases and the geometrical effects discussed previously overwhelm the favorable base-pairing free energy for the final few base pairs. This can be interpreted as an increase in the offset parameter $\delta_0$ with $f$, which helps to explain why $f_c$ is not perfectly linear in $\Delta G^\ddagger_{\tau_{\rm obs}}$. Unzipping lacks these kinds of geometrical effects; an equivalent graph of $f_c(N)$ against $\Delta G^\ddagger_{\tau_{\rm obs}}$ in Fig.\ref{fig:fc_as_fn_G}\,(b) does not show this apparent non-monotonic dependency of the fitted gradient with $N$. 

\begin{figure}
  \centering
  \includegraphics[width=.72\linewidth]{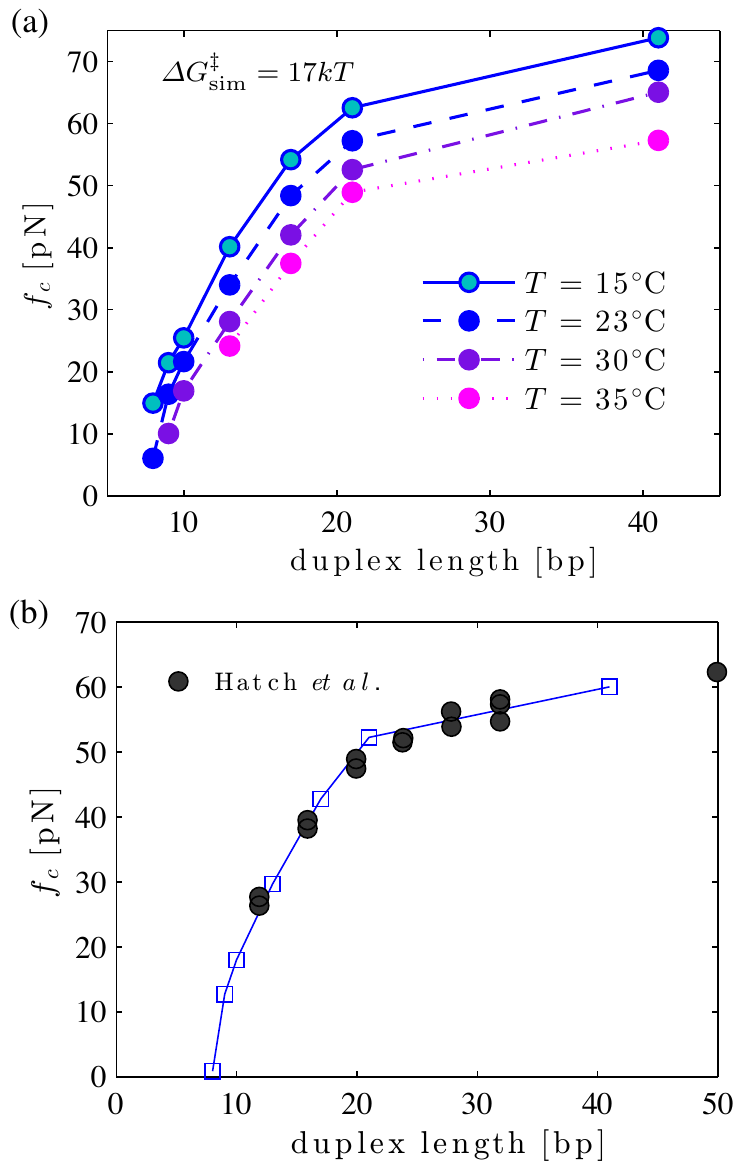}
 \caption{
 The consequences of temperature and base-pair stability. (a) Variation of $f_c(N)$ with temperature as predicted by oxDNA, defining $f_c(N)$ as the force required to reduce the free energy barrier, $\Delta G^\ddagger_{\tau_{\rm obs}}/kT$ to 17 (corresponding to \SI{10}{\kilo\cal\per\mole} at \SI{23}{\degreeCelsius}). (b) Fit of oxDNA model (squares) to the experimental data of Hatch {\it et al.}\:\cite{Hatch_pulling}. The fitting procedure generates a value for the rupture barrier height of $\Delta G^\ddagger_{\tau_{\rm obs}}=8.8$ \SI{}{\kilo\cal\per\mole} and includes a small rescaling of the temperature, for reasons described in the text. 
  }
  \label{fig:vary_T}
\end{figure}

The deGennes approach is based upon estimating the force at which each base pair is inherently unstable, so that rupture of the duplex occurs directly with each step being favorable. This criterion corresponds to the system ceasing to be metastable with a well-defined barrier. The toy model predicts that each base pair is inherently unstable when $f_c = \Delta G_{\rm bp} / \delta$; this force is $N$-independent. Thus any variation in $f_c(N)$ with $N$ as $\Delta G^\ddagger_{\tau_{\rm obs}} \rightarrow 0$ cannot be explained by the toy model: it must arise from considerations such as those of deGennes, or geometric factors such as those highlighted above that are absent in the toy model. We will now argue that, for oxDNA at least, these factors only influence $f_c(N)$ at very low $N$.

In Fig.\,\ref{fig:free-energy}\,(d) we show how $f_c(N)$ varies with $\Delta G^\ddagger_{\tau_{\rm obs}}$. The lowest value of $\Delta G^\ddagger_{\tau_{\rm obs}}$ (corresponding to the shortest time scale) that we consider is 3\,kcal\,mol$^{-1}$; smaller values make the unambiguous identification of maxima and minima in the free energy profile difficult, and indeed the assumptions inherent in our kinetic model are less robust. Even at this finite value of  $\Delta G^\ddagger_{\tau_{\rm obs}}$, however, $f_c$ depends strongly on $N$ only for $N \leq 10$; in the limit of  $\Delta G^\ddagger_{\tau_{\rm obs}} \rightarrow 0$, this dependence will be even further truncated. Indeed, the shape of $f_c(N)$ at $\Delta G^\ddagger_{\tau_{\rm obs}}=3$\,kcal\,mol$^{-1}$ still reflects the predictions of the toy model and its metastability-based arguments (cutting through $f_c=0$ at finite $N$, for example). The relatively short range of $N$ prior to the plateau as $\Delta G^\ddagger_{\tau_{\rm obs}} \rightarrow 0$ observed for oxDNA is consistent with the fact that geometric effects are only large for helices below a single pitch length, and that there is no physical reason why the $\chi$ parameter in Eq.\,\ref{eq:dG_prediction} should be much smaller than unity.

OxDNA allows exploration of the effect of temperature changes. In Fig.\,\ref{fig:vary_T}\,(a), we report $f_c(N)$ at a range of temperatures. 
Both $\Delta G^\ddagger$ and $kT$ in the exponent of Eq.\,\ref{eq:dGdagger} depend on $T$. Thus to compare critical forces at different temperatures for the same observation time, we find the force $f_c$ which gives $\Delta G^\ddagger_{\rm sim}/kT = 17$ (\SI{10}{\kilo\cal\per\mole} at \SI{23}{\degreeCelsius}, the standard temperature used in this study). We thereby include the full temperature dependence of the exponent in Eq.\,\ref{eq:Arrhenius}; any temperature dependence in the prefactor $k_0$ is neglected. Unsurprisingly, $f_c(N)$ is reduced by temperature, as base pairs become less stable and free energy barriers of a certain height slightly easier to climb. Consequently the plateau force $f_c(\infty)$ drops and the length of duplex below which $f_c=0$, $N_0$, rises. These changes with temperature are similar to those seen for the toy model when base pair stability is varied, as shown in Fig.\,\ref{fig:toymodel}\,(c) (note that in this case, the variation is performed at fixed $\Delta G^\ddagger  = \Delta G^\ddagger_{\tau_{\rm obs}}$).

The data of Hatch {\it et al.}\:were obtained at a lower salt concentration than those used to parameterize oxDNA to duplex formation in the absence of force ($\sim 0.15$\,M as opposed to 0.5\,M [Na$^+$]), implying weaker duplexes in experiment. Additionally, oxDNA is known to slightly overestimate (on the order of $10\%$) the critical force for the overstretching transition to ssDNA at 0.5\,M [Na$^+$]; this is essentially the critical shearing force for infinitely long strands. As a result, the direct fit of oxDNA to the experimental data (using only $\Delta G^\ddagger_{\tau_{\rm obs}}$ as a fitting variable) is only qualitative. To take these effects into account we fit to the data by shifting the temperature to $T = 35$\SI{}{\degreeCelsius}, a slight ($4\%$ absolute change) increase when compared to the $T = 23$\SI{}{\degreeCelsius}  of the experiments.   In addition to the temperature, we also need to fit $\Delta G^\ddagger_{\tau_{\rm obs}}$ which measures the effective barrier for rupture. Using these two fitting parameters, we find, as can be seen in Fig.\,\ref{fig:vary_T}\,(b),  very close agreement with the experiments of Hatch {\it et al.}\: The value of $\Delta G^\ddagger_{\tau_{\rm obs}}=8.8$ \SI{}{\kilo\cal\per\mole} is reasonable as it corresponds to approximately 6 or 7 base pairs \cite{SantaLucia2004}, a sensible number for spontaneous rupture on the time scale of seconds \cite{Morrison1993,Zhang_disp_2009, eyal_hairpin}. Whilst the temperature rescaling is crude, it is equivalent to a slight rescaling of the interaction strengths within oxDNA, producing a very similar model with slightly less stable base pairs. This oxDNA-like model, which still possesses mechanical and structural properties close to those of physical DNA, can quantitatively reproduce the experimental data. We therefore argue that the basic mechanism we propose, an activation based rupture of the duplex strands, can explain the experimental results.

\subsection{Mixed shearing and unzipping}

\begin{figure*}
  \centering
  \includegraphics[width=\linewidth]{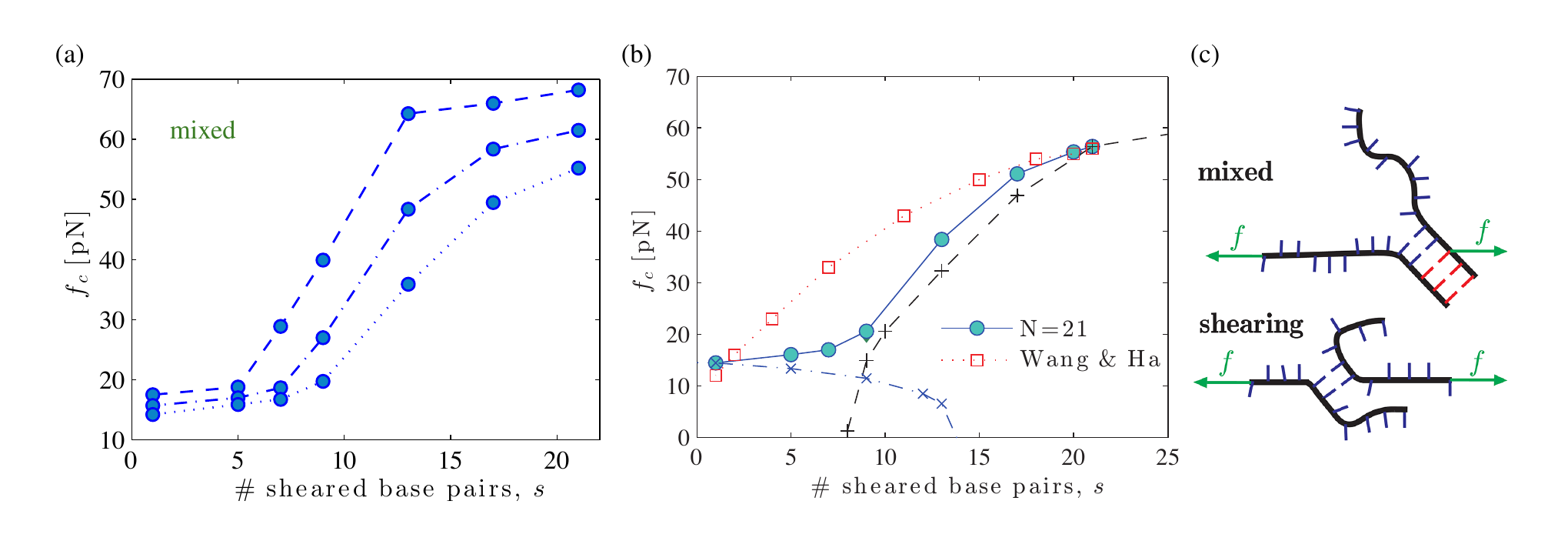}
 \caption{(a) Rupture force as a function of the number of sheared base pairs $s$, when a $21$-base-pair duplex is pulled in the mixed mode. $\Delta G^\ddagger_{\tau_{\rm obs}} \in \{5, 8, 11\}$ \SI{}{\kilo\cal\per\mole}, increasing from top to bottom. (b) Rupture force in the mixed mode as a function of sheared bases $s$ for a duplex of length $N=21$ (filled circles) ($\Delta G^\ddagger_{\tau_{\rm obs}} = 10.4$ \SI{}{\kilo\cal\per\mole} ). $(+)$ symbols show the corresponding rupture force for the pure shearing of duplex of length $s$ and $(\times)$ symbols represent pure unzipping rupture force of a duplex of length $N-s+1$. Open squares are Wang and Ha's estimate of the rupture force \cite{Wang_TGT}; their model is described in the text. (c) Schematic representation of the anchoring of the final sheared base pairs by the in the mixed mode by the base pairs that will eventually be unzipped; no such anchoring occurs for a pure shearing system.  
 }
  \label{fig:fc_N_dG}
\end{figure*}

Given the success of oxDNA in describing  the pure shearing mode shown in Fig.\,\ref{fig:modes}\,(a) we next turn to the mixed shearing and unzipping mode shown in Fig.\,\ref{fig:modes}\,(c). Wang and Ha have designed a tension gauge tether (TGT) to measure forces that involves applying stress in this ``mixed mode". In this setup, force is applied at the end of one strand and to any one base of the other strand, so that $s$ base pairs are sheared and $N-s+1$ base pairs are unzipped. Note that the total appears to be $N+1$: this is because there is no difference between unzipping and shearing of a single base pair; for the analysis below it is helpful to define both quantities to include this ambiguous base pair. OxDNA allows us to explore the physics of this more complex system, using the same assumption that $f_c$ is the force required to reduce $\Delta G^\ddagger_{\tau_{\rm obs}}$ to a specific value. Fig.~\ref{fig:fc_N_dG}\,(a) shows the rupture force for a duplex of length $N=21$\,bp that is pulled in a mixed mode as a function of the number of sheared base pairs $s$. We see a characteristic sigmoidal curve for $f^{\rm mixed}_c(N,s)$ at fixed $N$. $f^{\rm mixed}_c(N,N)$ is identical to our pure shearing data in Fig.~\ref{fig:free-energy}\,(d) and depends on $\Delta G^\ddagger_{\tau_c}$ as expected. $f^{\rm mixed}_c(N,1)$ is identical to the pure unzipping data.

The sigmoidal curves therefore interpolates between the unzipping and shearing limits. A lower bound on the critical force is given by
\begin{equation}
f^{\rm mixed}_c(N,s) \geq {\rm max} \left( f^{\rm shear}_c(s), f^{\rm unzip}_c(N) \right),
\end{equation}
as is clear from Fig.\,\ref{fig:fc_N_dG}\,(b). The bound follows from two arguments. Firstly, the system can be viewed as a sheared duplex of $s$ base pairs, with additional base pairs that must also be unzipped. Although these additional base pairs might not substantially stabilize the TGT at high force, they cannot act to destabilize it either; thus $f^{\rm mixed}_c(N,s) \geq f^{\rm shear}_c(s)$. Secondly, sheared base pairs are always more stable than base pairs subject to an unzipping force (Fig.\,\ref{fig:free-energy}\,(c)), due to the far greater increase in extension upon rupture in the second case. Therefore the system must be at least as stable as an $N$-base-pair duplex subject to pure unzipping; $f^{\rm mixed}_c(N,s) \geq f^{\rm unzip}_c(N)$.

At large forces, the unzipped base pairs  contribute almost nothing to the barrier to dissociation and hence do not strongly affect the stability of the TGT, because at forces $f> f^{\rm unzip}_c(\infty)$ each base pair subject to an unzipping force is inherently unstable. Thus at large $s$, $f^{\rm mixed}_c(N,s)$ follows $f^{\rm shear}_c(s)$ very closely as is evident in Fig.\,\ref{fig:fc_N_dG}\,(b). On closer inspection, we note that in fact $f^{\rm mixed}_c(N,s)$ is slightly greater than $f^{\rm shear}_c(s)$ for $s<N$. This discrepancy arises because in a purely sheared system, the base pairs present in the transition state can be anywhere within the duplex. In the TGT however, the unzipped base pairs at the end of the duplex anchor the final sheared base pairs (as illustrated schematically in Fig.\,\ref{fig:fc_N_dG}\,(c)). This anchoring decreases the entropy of the transition state, and thus increases the barrier to TGT rupture slightly.           

Together, the lower bound and the insight that $f^{\rm mixed}_c(N,s) \approx f^{\rm shear}_c(s)$ for $ f^{\rm mixed}_c(N,s) > f^{\rm unzip}_c(\infty)$ imply the sigmoidal curve for $f^{\rm mixed}_c(N,s)$ at fixed $N$. At $s = N$, $f^{\rm mixed}_c(N,s) = f^{\rm shear}_c(N)$. As $s$ decreases, the curve $f^{\rm mixed}_c(N,s)$ approximately follows $f^{\rm shear}_c(s)$, which has the characteristic shape discussed previously and $f^{\rm mixed}_c(N,s)$ drops towards $0$ increasingly rapidly. Eventually, however, the lower bound $f^{\rm mixed}_c(N,s) \geq f^{\rm unzip}_c(N)$ comes into play, forcing the curve to plateau at low $s$.       

Wang and Ha calibrated their TGT by assuming that $f^{\rm mixed}_c(N,s) = f^{\rm shear}_c(s)$ \cite{Wang_TGT}, with $f^{\rm shear}_c(s)$ given by Eq.\,\ref{eq:hatch_prediction} (the formula of Hatch {\it et al.}\:\cite{Hatch_pulling}). They used the parameterization of Hatch {\it et al.}, but set $N_{\rm open} =0$ rather than $N_{\rm open} =7$; due to this choice, $f^{\rm shear}_c(s)\sim 10 $\,pN at $s=1$, rather than 0 (or 3.9\,pN, as would be predicted by the original deGennes formula \cite{deGennes_ladder_model}). Their resultant calibration curve is shown in Fig.\,\ref{fig:fc_N_dG}\,(b). Note that the shape is quite different from that predicted by oxDNA;  in particular it has no inflection. The Wang and Ha curve predicts a physically reasonable $f_c$ for $s \approx N$ by design, and, due to the choice of using the Hatch parameterization of Eq.\,\ref{eq:hatch_prediction} but neglecting $N_{\rm open}$, predicts a roughly reasonable value for $s=1$. In this picture, however, the finite $f_c$ at $s=1$ is due to the robustness of a single sheared base pair rather than a duplex of length $N$ that is subject to unzipping. Our results suggest that given agreement at $s=1$ and $s=N$, the critical forces predicted by Wang and Ha's calibration approach would be far too high at intermediate values of $s$. Furthermore, we argue that $f^{\rm mixed}_c(N,s)$ depends on the time scale of observation; a consideration that does not appear in the Wang and Ha calibration curve.

\section{Conclusion}   
In this article we argue that shearing of short DNA duplexes should be understood as an activated process, in which a metastable duplex dissociates into single strands. Whether or not a certain force is large enough to cause dissociation within an experiment therefore depends fundamentally on the  observation time $\tau_{\rm obs}$; in principle, all duplexes subject to a constant force will dissociate permanently given enough time.

We used an extremely simple toy model to highlight the basic predictions of such an understanding. For a given $\tau_{\rm obs}$, we must apply a large enough force to reduce the free-energy barrier opposing dissociation, $\Delta G^\ddagger$, to $\Delta G_{\tau_{\rm obs}}^\ddagger$. Since more base pairs must be disrupted in longer duplexes to reach the transition state, the force-influenced stability per base pair must be less to achieve $\Delta G^\ddagger=\Delta G_{\tau_{\rm obs}}^\ddagger$; hence the critical force $f_c$ increases with duplex length $N$. Once the force is large enough to reduce the free-energy gain per base pair to zero, however, even the longest duplexes will have $\Delta G^\ddagger<\Delta G_{\tau_{\rm obs}}^\ddagger$. This introduces a maximum $f_c$, leading to a plateau in $f_c(N)$ for long duplexes. Importantly, this simple picture naturally predicts that  $f_c(N)$ tends to zero at finite $N$. This overall behavior is consistent with experimental observations.

To go beyond the simple model, we have used oxDNA, a coarse-grained model of DNA, to obtain more physically realistic force-dependent free-energy barriers. OxDNA reproduces the basic behavior predicted by the toy model, despite the more complicated free-energy profiles and force-extension properties. In particular, oxDNA also predicts that $f_c(N)$ tends to zero at finite $N$, and that $f_c(N)$ includes an approximately logarithmic dependence on the observation time. OxDNA is itself a simplified model, but the key aspects of DNA relevant to this study (force-extension properties, thermal stability, geometry) are known to be physically reasonable \cite{tom_model_jcp, tom_thesis}. The generic results are therefore likely to be robust; a detailed investigation into the relationship between rates and free-energy barriers for sheared oxDNA molecules, incorporating possible mis-aligned base pairs, is ongoing \cite{rupture_jcp}. We have also demonstrated that, by slightly adjusting the interaction strengths in oxDNA, 
in part to take into account the fact that it was parameterized at a different salt concentration, 
we can fit the available experimental data quantitatively, supporting the essence of our claims. 

Previous work has contended that the shape of $f_c(N)$ arises because a shorter duplex is immediately unstable to rupture (rather than metastable with a short enough life time) at a lower force than longer duplexes \cite{Hatch_pulling, deGennes_ladder_model, Chakrabarti_shearing,Prakash2011,Mishra2011}. We see evidence that shorter duplexes are indeed unstable at lower forces in our data for oxDNA, but we argue that this is not the primary factor in determining the shape of $f_c(N)$. Conceptually, it is clear that any theory that describes whether a molecular system undergoes change on a time scale of seconds or more should allow for metastability. In terms of concrete experimental evidence, it is known that duplexes of a few base pairs will spontaneously dissociate in the absence of force on a short time scale \cite{Morrison1993,Zhang_disp_2009, eyal_hairpin}. This is consistent with the data of Hatch {\it et al.}\:that show $f_c(N)$ tending to zero at finite $N$ \cite{Hatch_pulling}. This behavior does not naturally arise in models that predict $f_c(N)$ based on absolute instability rather than metastability. Further evidence against the deGennes' model arises from the observation that Hatch {\it et al.}\:needed to use unphysical values for the ratio of the spring constants between stacked neighbors in the same strand and hydrogen-bonded base pairs on opposite strands in order for the  theory to fit their data.

Further tests of the mechanisms that determine the critical force could be explored by experiments in which the observational time scale is varied. Hatch {\it et al.}\:did consider time scales between 1\,s and 3\,s, but given the approximately logarithmic variation expected in $f_c$ with $\tau_{\rm obs}$, it is unsurprising that systematic effects were not visible above experimental noise. The approach presented in this work predicts a change in $f_c(N)$ with $\tau_{\rm obs}$; previous arguments based on absolute instability do not. Data from a wider range of duplex lengths $N$ would help to differentiate between the typical shapes of $f_c(N)$ predicted by the two curves. We also make predictions for the temperature dependence of $f_c(N)$, which could be tested in experiments.

We have also used oxDNA to predict $f^{\rm mixed}_c(N,s)$ for a mixed system in which $s$ of the $N$ base pairs are sheared, and the remainder are subject to an unzipping force. Such a system is the basis of the interesting tension gauge tether (TGT) proposed by Wang and Ha \cite{Wang_TGT} for measuring biomolecular forces. We argue that the shape of $f^{\rm mixed}_c(N,s)$ is very different from previously supposed, showing a characteristic sigmoidal shape as $s$ is increased at fixed $N$. At small $s$, $f^{\rm mixed}_c(N,s)$ approaches the pure unzipping force $f^{\rm unzip}_c(N)$, but approximates the critical force for shearing an $s$-base duplex, $f^{\rm shear}_c(s)$, at larger $s$. Not only do we predict that careful calibration of the TGT will reveal this complex $f^{\rm mixed}_c(N,s)$, we also claim that the time scale of experimental observation will influence $f^{\rm mixed}_c(N,s)$, in analogy with the pure shearing system. Thus quantitative use of the TGT will require extensive calibration.

\begin{acknowledgments}
 The computational resources of the PolyHub virtual organization are gratefully acknowledged. M.M. was supported by the Swiss National Science Foundation (Grant No. PBEZP2-145981). T.E.O. acknowledges funding from University College, Oxford. The authors are also grateful to the EPSRC for financial support. M.M. thanks Flavio Romano for helpful discussions.
\end{acknowledgments}


%

\end{document}